\documentclass[pla,12pt]{elsarticle}

\usepackage{amsmath,amssymb,graphicx, xcolor}

\begin{document}
	
	\begin{frontmatter}
		
		\title{Mass Oscillations in Superconducting Junctions for Gravitational Wave Emission and Detection}
		
		\author{Victor Atanasov} 
		
		\affiliation{organization={Faculty of Physics, Sofia University},
			addressline={5 blvd J. Bourchier}, 
			city={Sofia},
			postcode={1164}, 
			state={},
			country={Bulgaria}}

		\author{Avadh Saxena} 
		
		\affiliation{organization={Theoretical Division and Center for Nonlinear Studies, Los Alamos National Laboratory},
			city={Los Alamos},
			postcode={87545}, 
			state={NM},
			country={USA}}

		\begin{abstract}
			We revisit the nonlinear superconducting junction dynamics in order to provide evidence that the time-dependent current density in the junction is related to an oscillating  charge and mass density in addition to a variable velocity. As a result, the superconducting tunnel junction emerges as a solid state device capable of producing rapid charge and mass oscillations inaccessible in classical contexts. Rapidity is required for gravitational wave emission when small masses are involved in the emission process. We provide designs for an emitting and a detecting device based on generating and capturing mass oscillations with a non-zero quadrupole moment component. Finally, we suggest that the smallness of the  Einstein gravitational coupling constant can be fully compensated by the largeness of the quantum mechanical term $e^4 \hbar^{-6}$ manifested in the suggested set ups.
		\end{abstract}

		\begin{highlights}
			\item An exact solution to the nonlinear dynamics of the superconducting tunnel junction is provided.
			\item Charge and mass oscillations in a SC tunnel junction are shown to exist.
			\item An oscillator model for a SC tunnel junction is constructed.
			\item A gravitational wave emission from a quadrupole SC tunnel junction device is discussed.
			\item Detection of gravitational waves with quadrupole SC tunnel junction devices is analyzed. 
		\end{highlights}

	\end{frontmatter}

\section{Introduction}

The solid state has the potential to bridge our gravitational technology from the purely observational state to the one where we can actually manipulate the space-time itself. When it comes to emission, it is the kinetic energy and a non-vanishing time dependent quadrupole mass moment of matter that drives this classical process of geometric field generation \cite{LL}. The detection is multifaceted and as we will demonstrate here not restricted to classical physics only \cite{GW1, GW2, GW3, GW4}.  

Laser Interferometer Gravitational-Wave Observatory (LIGO) interferometry experiments observe the traveling strain in space-time generated by collisions of massive astrophysical objects such as neutron stars and black holes \cite{adh, ligo}. The strain carried by the generated gravitational wave (GW) is presently inaccessible for the solid state, mainly due to it being negligibly small (less than the size of a single nucleus in the case of LIGO's kilometer size).

The laboratory generation and tandem detection of GW has been the focus of now abandoned research \cite{weber, Forward, Gertsenshtein, Halpren, Romero}. The most promising solid state effect for GW physics is the piezoelectric effect, where the strain, that is mass displacement as well, is converted into a measurable response, such as a voltage drop, in effect a strain-to-voltage reversible conversion. Unfortunately, the piezoelectric effect is not sufficiently big for the task of gravitational wave detection. The reverse, that is a periodic voltage which produces deformation in the piezoelectric material, is also not a viable option for gravitational wave emission, due to (i) the problem of energizing a piezoelectric array and (ii) the frequency of such a mass oscillation is typically low and therefore produces undetectable GW emission power \cite{pz1}. The piezoelectric effect is somewhat classical in nature and we can never hope to find fast enough mass oscillations for GW emission outside the quantum mechanical realm.

An additional confirmation that the classical realm, including Micro Electro Mechanical Systems (MEMS) actuators and accelerometers (sensors), is incapable of presenting a context for mass oscillations relevant to GW emission or detection can be found in Refs. \cite{mems2,mems3, mems4, mems5}.  

One of the fastest quantum processes outside the molecular and atomic structure is to be found in the solid state. More specifically, the AC or DC tunneling dynamics across a superconducting junction \cite{bj1, bj2}. The characteristic number setting the scale of the rapidity is the Josephson constant: a DC voltage of 1 mV across the junction causes an oscillation frequency of the charged and massive superconducting condensate of about 483 GHz \cite{FF, JER}.

A common misunderstanding of the super-current dynamics exists. Naturally, the current  
is a product of the superconducting pair density $\rho$ and its velocity $v$ (in the rest frame
of the lattice), that is $j=\rho v$. It is assumed that the Josephson current oscillations do not automatically imply charge (mass) density oscillations, but it is rather the velocity that oscillates $j(t)=\rho v(t)$.

Here we ask the question: {\it Do superconducting junction current oscillations imply
mass oscillations as well?} Provided the charge density component of the current density does not remain constant, that is $j(t)=\rho(t) v(t)$, the answer is in the affirmative. Otherwise, the oscillation actually affects the drift velocity of the carriers which is the standard interpretation of the dynamics across the tunnel junction \cite{feynman, lik}.

The oscillating superconducting condensate charge density implies the tunneling of the charge carriers, i.e. Cooper pairs which are well explored quasi-particles with material dependent effective mass. The typical range of the effective mass for these carriers is between two and five (vacuum) electron masses \cite{CP1, CP2, CP3, CP4}. Naturally, the gravitational response is associated with the effective mass (the total relativistic mass-energy of the system of nucleons) as is demonstrated in nuclear physics (the atomic mass is not equal to the mass number)  and the Stewart-Tolman effect, which points to the effective mass, in the condensed matter context, participating in inertial phenomena \cite{ST1, ST1_2,  ST2}. The effect is material specific, which is a confirmation that the effective mass of the quasi-particle is what is taking part in the inertial response. As a result, the oscillation (motion under variable acceleration) of the superconducting condensate is associated with an oscillating mass as well.

Note, the issue is rather important when it comes to quantum computation \cite{Qbit1, Qbit2}. At present, the standard qubit in commercial applications of quantum computation (Google, IBM, D-Wave, Rigetti, etc.) is a superconducting Josephson junction. The answer to the above posited question can provide means to further reduce the rate of decoherence of the quantum states and therefore can provide unforeseen benefits in error correction, ergo enlargement of the qubit base involved in the computation \cite{Qbit3}.

The Josephson junction which we will call superconducting tunnel junction in this paper is treated as a two level quantum system, which is subject to an external driving force and the law of conservation of charge. Therefore, the time dependence of the charge densities ($\rho_j$ for $j=1,2$) in the two sides of the juction should be related by: $\dot{\rho}_1=-\dot{\rho}_2 \propto \sqrt{\rho_1 \rho_2} \sin{\phi},$ where $\phi$ is the phase difference of the condensate's wave function across the junction. The relation is a statement that if current flows from one of the sides, it charges up the other side which changes its electric potential. It is precisely this behavior that leads to high frequency oscillations. The frequency of these oscillations is determined by the Josephson constant $K_J = 2e/h$, which is 483.598  THz V$^{-1}$. 

However, when the junction is connected to a source of an electromotive force, i.e. a battery (constant electric potential $V$), currents will flow to reduce the potential difference across the junction. When the currents from the battery are included in the Josephson junction description, standard solution dictates that charge densities do not change and there should be no mass oscillation. This is the famous $\rho_1= \rho_2= \rho_0$ solution, where $\rho_0$ is a constant, i.e. the standard solution.
 
The standard solution is correct provided $\sin{\phi}=0,$ which happens  either by (i) $\phi \propto t$ and is {\it vanishing only as a mean value} (rapid oscillations which means that the average amplitude is assumed zero), (ii) $\phi = {\rm n} \pi$ for ${\rm n} \in\mathbb{Z},$ which is a form of an unjustified quantization condition, or (iii) $V=0,$ that is the static mode. 

Note, in general, the system of equations that describes the dynamics in the superconducting tunnel junction is highly nonlinear. The standard solution and description are approximate. The DC mode is understood as $\rho_1= \rho_2=\rho_0$ and $\phi=2eVt/\hbar,$ which leads to
\begin{eqnarray}
\dot{\rho}_1=-\dot{\rho}_2 \propto \sin{2eVt/\hbar} \neq 0 \,, 
\end{eqnarray}
a stark contradiction due to the approximate character of the solution. 

In this paper we resolve this contradiction and present evidence, that charge and mass rapidly oscillate in the superconducting tunnel junction. This is done in sections ~\ref{sec.2lev} and \ref{sec.Smallpar}. Section \ref{sec.emission} is dedicated to the practical use of such rapid mass oscillations, namely the artificial generation of gravitational waves. 
In the subsequent section \ref{sec.detection}, we propose a solid state device, which can serve as a gravitational wave detector, that is a curvature-to-frequency converter with much higher sensitivity than the piezoelectric effect. The device's current-voltage characteristic is explored in the subsequent section. The paper ends with a conclusion in section 8.


\section{A two-level quantum system}\label{sec.2lev}

The weakly coupled superconducting tunnel junction can effectively be modeled as a two-level quantum system \cite{feynman, lik}. The rate of change of the superconducting pairs' wavefunction on one side of the junction depends on the instanteneous values of the wavefunction on both sides, subscript 1 and 2:  
\begin{subequations}
\begin{eqnarray}\label{2level}
i \hbar \partial_t \psi_1 &=&  \frac{qV}{2}  \psi_1 + K \psi_2 \,, \\ 
i \hbar \partial_t \psi_2 &=& K^{\ast} \psi_1 - \frac{qV}{2} \psi_2\,. 
\end{eqnarray} 
\end{subequations}
Here $V$ is the applied voltage bias which measures the difference in the superconducting pairs' self energies $U_i$ (for $i=1,2$) across the insulating layer $\lvert U_2 - U_1 \rvert = qV$. The charge $q=2e$ is the charge of the Cooper pairs. The coupling energy coefficient $K$ is a real-valued constant for the  magnetic-free case ($K=K_0$) and is accompanied by a factor which depends on an integral over the magnetic vector potential across the junction in the general case:
\begin{eqnarray}
K (\vec{A}) = K_0 e^{i \kappa } \,, \quad \kappa=\frac{2e}{\hbar} \int_{1}^{2} \vec{A} \cdot d \vec{r} \,. 
\end{eqnarray} 

Using the trigonometric representation for the wavefunction $\psi_{j}=\rho_{j} e^{i \nu_{j}},$ where $i=1,2$ and separating the real and imaginary parts leads to the nonlinear system of equations
\begin{subequations}
\begin{eqnarray}\label{rho1}
\partial_t \rho_1 &=&  \frac{2K_0}{\hbar} \sqrt{\rho_1  \rho_2} \sin{\phi} \,,  \\\label{rho2}
\partial_t \rho_2 &=&  - \frac{2K_0}{\hbar} \sqrt{\rho_1  \rho_2} \sin{\phi} \,, \\\label{nu1}
\partial_t \nu_1 &=& -\frac{K_0}{\hbar}\sqrt{ \frac{\rho_2}{\rho_1}  } \cos{\phi} - \frac{qV}{2\hbar} \,, \\\label{nu2}
\partial_t \nu_2 &=& -\frac{K_0}{\hbar}\sqrt{ \frac{\rho_1}{\rho_2}  } \cos{\phi} + \frac{qV}{2\hbar} \,. 
\end{eqnarray} 
\end{subequations}
Here $\phi=\theta + \kappa$ and $\theta=\nu_2 - \nu_1$. From (\ref{rho1}) and (\ref{rho2}) we obtain
\begin{eqnarray}\label{rho1=-rho2}
\partial_t \rho_1 = - \partial_t \rho_2 ,
\end{eqnarray} 
which is the mathematical expression for the law of conservation of charge and has the obvious solution $\int \partial_t \rho_1 dt = - \int \partial_t \rho_2 dt$, that is
\begin{eqnarray}\label{sol1_rho1}
\rho_2(t) =2\rho_0 -\rho_1(t).
\end{eqnarray} 
Here $\rho_0$ is a constant, which is interpreted as the equilibrium charge density. Therefore, we insert (\ref{sol1_rho1}) in (\ref{rho1}) to obtain
\begin{eqnarray}\label{}
\partial_t \rho_1 =  \frac{2K_0}{\hbar} \sqrt{\rho_1 \left( 2\rho_0-\rho_1 \right)} \sin{\phi}.   
\end{eqnarray} 
Introducing a dimensionless variable $u=\rho_1/2\rho_0,$ the above equation can be cast into
\begin{eqnarray*}\label{}
\frac{d u}{\sqrt{u -u^2} } =  \frac{2K_0}{\hbar} \sin{\phi} \, dt \,, 
\end{eqnarray*} 
which is easily integrated \cite{dwight}
\begin{eqnarray*}\label{}
\sin^{-1} \left( 1-2u\right) = \theta_0 - \frac{2K_0}{\hbar} \int  \sin{\phi} \, dt.
\end{eqnarray*} 
Here $\theta_0$ is an integration constant. Finally,
\begin{subequations}
\begin{eqnarray}\label{sol_rho1}
 \rho_1 (t)&=& \rho_0 \left[ 1 -  \sin{f(t) } \right] \,, \\\label{sol_rho2}
\rho_2 (t)&=& \rho_0 \left[ 1 +  \sin{f(t) } \right] \,, 
\end{eqnarray} 
\end{subequations}
where
\begin{eqnarray}\label{f(t)}
f(t) &=& \theta_0  - \frac{2K_0}{\hbar} \int  \sin{\phi}\, dt .
\end{eqnarray}

Regardless of the specific temporal evolution of $f(t),$ the mean value
\begin{eqnarray}
\langle \rho_1 \rangle_t =\langle \rho_2 \rangle_t =  \rho_0 
\end{eqnarray} 
coincides with the standard result, commonly misunderstood as exact. The exact solution, (\ref{sol_rho1}) and (\ref{sol_rho2}), clearly exhibits behavior which can be interpreted as charge and more interestingly {\it mass density oscillation}. The mass density oscillation comes as an epiphenomenon of the charges (electrons, superconducting pairs) being massive particles, including in the sense of effective mass in the solid state.

\section{Small parameter solution: oscillator equation}\label{sec.Smallpar}

The exact solution presented in the previous section clearly exhibits the sought after mass oscillation. However, it depends on the temporal behavior of the phase $\phi(t)$. The equation describing the evolutionary behavior of the phase $\phi$ is highly nonlinear. We are not able to provide a general analytical solution, but would once again confirm the oscillatory behavior of the charge and mass density across the superconducting junction.

Suppose $\epsilon(t) \ll \rho_0$, then a natural small parameter $x(t)=\epsilon /\rho_0 \ll 1$ emerges. Preserving the form of the exact solution (\ref{sol_rho1}) and (\ref{sol_rho2}), we can assume that 
\begin{subequations}
\begin{eqnarray}
 \rho_1 (t)= \rho_0 - \epsilon(t) \,, \\
\rho_2 (t)= \rho_0 + \epsilon(t) \,,
\end{eqnarray} 
\end{subequations}
is a legitimate type of solution.  Inserting this ansatz in the original system (\ref{rho1}) we obtain
\begin{eqnarray}\label{eq:x}
\nonumber \partial_t x &=&  - a \sqrt{1-x^2 } \sin{\phi} \\
&=& - a \sin{\phi} + o(x^2),
\end{eqnarray}
where $a={2K_0 }/{\hbar}$. Note that we have expanded the r.h.s. of the equation over the small parameter and kept the first order terms only. The second equation for the phase is further simplified if we assume that $\kappa$ is a constant, that is a vanishing or orthogonal to the trajectory of the superconducting pairs vector potential, $\vec{A}$:
\begin{eqnarray}\label{eq:phi}
\partial_t \phi =   a x(t)  \cos{\phi}  + b,
\end{eqnarray}
where $b={qV}/{\hbar}.$ Let us now differentiate equation (\ref{eq:x}) with respect to time and substitute the derivative of the phase from equation (\ref{eq:phi}):
\begin{eqnarray}\label{}
\partial_t^2 x + (a \cos{\phi})^2 \;  x(t) = - ab \cos{\phi}. 
\end{eqnarray}
The structure of this equation is rather obvious
\begin{eqnarray}\label{}
\partial_t^2 x + \omega_0^2(t) \;  x(t) = f(t).
\end{eqnarray}
 It is a driven oscillator with resonant frequency $\omega_0(t)=|a \cos{\phi}|$ and a driving force $f(t)=-b \omega_0(t)$. Note, the driving force is composed of two terms. One is $b \propto V(t)$ proportional to the applied voltage across the superconducting junction. As is well known the applied voltage can be either DC, therefore $b$ will set the fixed amplitude of the driving force, or AC in which case $b(t)$ will set the time dependent amplitude, whose frequency, if lower than $\omega_0$, will serve as an envelope modulation of the driving force. 

More interestingly, the second term in the driving force is exactly equal to the resonant frequency of the oscillator, therefore the oscillator is being {\it driven always resonantly}, namely $x(t)$ is a periodic function with frequency $\omega_0.$ 
This suggests that charge and mass oscillations with frequency $\omega_0$ at superconducting junctions are admissible solutions.

Now we turn to an estimate of the squared resonant frequency using its mean value. 
\begin{eqnarray}\label{}
\langle \omega_0^2 \rangle_t   = a^2 \langle \cos^2{\phi} \rangle_t = a^2/2=\frac{2 K_0^2 }{\hbar^2}.
\end{eqnarray} 
Naturally, the resonant frequency is proportional to the coupling energy over Planck's constant. This is a very large frequency due to $\hbar \propto 10^{-34}$ being in the denominator.

\section{Mass oscillations for gravitational wave emission}\label{sec.emission}

Gravitational-wave sources are mainly time-dependent quadrupole mass moments \cite{LL}. The traceless $D_{\alpha \alpha}=0$ mass quadrupole moment tensor is given by $D_{\alpha \beta}=\int \rho (3 x_{\alpha} x_{\beta}-r^2 \delta_{\alpha \beta}) dV$. Here $\rho$ represents the mass density which in the case of a superconducting tunnel junction is proportional to the charge density. A restriction to axial symmetry (with respect to the $z$-axis) diagonalizes the tensor in a manner that $D_{xx}=D_{yy}=-D_{zz}/2$. Furthermore, if mass moves along the $z$-axis only (imposing one-dimensional motion by setting $x=0$ and $y=0$), then 
\begin{eqnarray}\label{}
D_{zz}=2 \int \rho z^2 dV \,, 
\end{eqnarray} 
and as a result $\dddot{D}_{\alpha \beta}^{\quad 2} = 3\dddot{D}_{zz}^{\; 2}/2 $, where $\dddot{D}$ represents the third derivative with respect to time.

The power of gravitational wave emission is governed by
\begin{eqnarray}
P &=& \frac{G}{30 c^5}  \dddot{D}_{zz}^{\quad 2}.
\end{eqnarray} 
Note the smallness of the coupling constant $G/c^5 \approx 3 \times 10^{-53}$ s$^3$ kg$^{-1}$ m$^{-2}$, where $G$ is Newton's gravitational constant and $c$ is the speed of light in vacuum. The standard argument for circumventing the smallness of the constant in gravitational wave production is the requirement of astronomically large masses which can produce substantially large quadrupole mass tensor. Here and elsewhere \cite{VA}, we convey an argument in favor of a high frequency process which can substitute the necessity for large masses in gravitational wave production. Such a high frequency process can be initiated in a quantum mechanical setting, namely the oscillation of the superconducting condensate between two separated superconducting regions of a tunnel junction.

Next, suppose the current density alternating between the sides of the junction device sets the law of motion for the superconducting pairs, that is the same oscillation as the one given in equations (\ref{sol_rho1}, \ref{sol_rho2}) and (\ref{f(t)}):
\begin{eqnarray}\label{z(t)}
z(t)=A_z  \sin{f(t)}.
\end{eqnarray}
Here $A_z$ is the amplitude of the oscillation. It is probably equal to a few times the thickness of the insulating barrier in the junction. Note, $\rho(t)$ can be either $\rho=\rho_1(t)$ or $\rho=\rho_2(t)$ without a difference for the end result. We will use the notation $\int \rho_0 dV = M$, where $M$ is the total mass of the Cooper pairs sloshing between the sides of the junction. Here $\rho_0$ will be the equilibrium mass density which is proportional to the charge density as discussed above. Finally:
\begin{eqnarray}\label{}
\nonumber D_{zz}&=&2 A_z^2 \int \rho_0 (1+\sin{f}) \sin^2{f} dV \\
&& = 2 A_z^2 M (1+\sin{f}) \sin^2{f} \,, 
\end{eqnarray} 
which has the following third order derivative
\begin{eqnarray}\label{}
\nonumber &&\dddot{D}_{zz} = \frac{ A_z^2 M}{2} \times \\
\nonumber&& \left[ \dddot{f} \left(4 \sin{2f}+3\cos{f} - 3 \cos{3f} \right) \right.\\
\nonumber  && +3 \ddot{f} \dot{f} \left( -3 \sin{f} + 9 \sin{3f}+8\cos{2f} \right)  \\
\nonumber  && + \dot{f}^3 \left. \left( -16\sin{2f} -3 \cos{f} + 27 \cos{3f} \right) \right]  \,. 
\end{eqnarray} 
Substituting the time derivatives of $f$ with their approximate corresponding values, we obtain an expression which can be further reduced by assuming that $qV \gg K_0$, that is the applied electrical energy across the junction is far greater than the coupling energy: 
\begin{eqnarray}\label{}
\nonumber \dddot{D}_{zz} &\approx & \frac{ A_z^2 M q^2 V^2 K_0}{\hbar^3}  \left[  \sin{\phi} \left(4 \sin{2f}   \right. \right. \\
&&\quad \left.  \left. + 3\cos{f} - 3 \cos{3f} \right) \right] \,. 
\end{eqnarray}

\begin{figure}[t]
	\centering
	\includegraphics[scale=0.35]{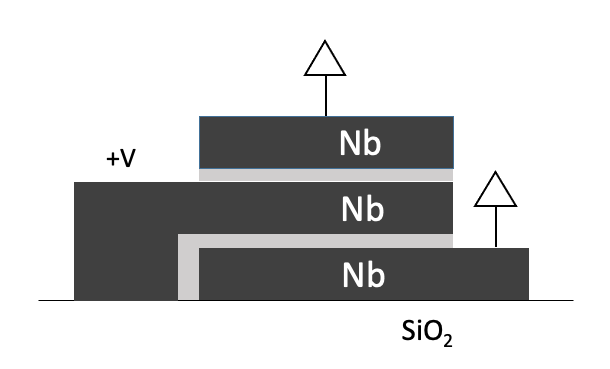}
	\caption{A gravitational wave emitter can be micro-fabricated on an insulating substrate by sequential evaporation/deposition of niobium metal. The insulating layers (thin light gray separations) are essential for the operation of the tunnel junction.}
	\label{fig1}
\end{figure}

Now, the averaged power of gravitational wave emission comes out to be as 
\begin{eqnarray}
\nonumber P 
&\approx& \left( \frac{Ge^4}{c^5 \hbar^6} \right) 5 A_z^4 M^2  V^4 K_0^2,  
\end{eqnarray} 
where we have used the fact that the charge of the Cooper pairs $q=2e$ is twice the charge of the electron. Interestingly, the smallness of the coupling constant $G/c^5$ is no longer an issue since $\hbar^6$ appears in the denominator, that is the coupling constant for this quantum process is given by
\begin{eqnarray}
\frac{Ge^4}{c^5 \hbar^6}  \approx 6.5 \times 10^{128}\quad  {\rm C^{-3} s^{-5} V^{-7}  }.
\end{eqnarray} 
The structure of this newly emerged coupling constant reveals an interesting behavior at the superconducting tunnel junction, namely {\it the dependence on the charge (electromagnetic energy) is stronger than the dependence on the mass}. In other words, this set up channels electromagnetic energy into the geometric (gravitational) field with greater ease than simply converting mass into curvature (the classical Einsteinean mechanism). We suppose that this is possible due to the quantum nature of the superconducting tunnel junction.

The minimum design capable of producing a non-vanishing time dependent quadrupole mass moment consists of two superconducting tunnel junctions on a line and is discussed in \cite{VA}. The two ends of this line are grounded and the device is driven by an AC/DC voltage applied at the center. Such a device can be micro-fabricated and one possible such fabrication based on niobium metal (with the superconducting transition temperature $T_C = 9.26$ K) is depicted in Figure \ref{fig1}.

\section{Charge oscillations for gravitational wave detection}\label{sec.detection}

At present, we are not aware whether the signatures of mass oscillations in the solid state are measurable, beyond the charge redistribution. However, it is rather important to provide a way in which one can ascertain whether the mass oscillations are due to the gravity waves and not of any other origin. 

\begin{figure}[t]
	\centering
	\includegraphics[scale=0.25]{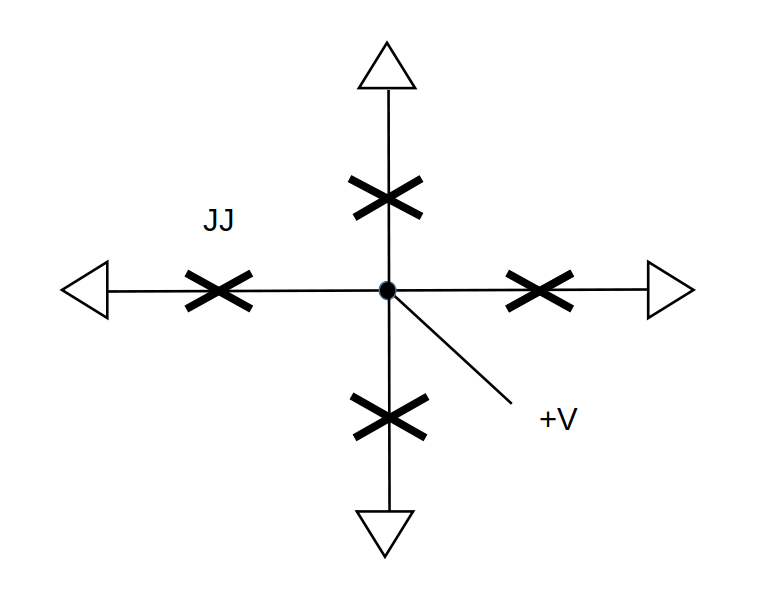}
	\caption{GW detector based on the voltage-to-frequency conversion at a superconducting tunnel junction. The passing GW causes mass and therefore charge displacement, which leads to an oscillating current between the center point and the ground. The frequency of oscillation is a function of the displaced charge.}
	\label{fig2}
\end{figure}

Here we refer back to the main feature of gravitational waves: they are the product of oscillating quadrupole mass distributions and therefore cause such types of strain in the fabric of space-time. As a result, if the detection device is sensitive to quadrupole mass oscillations we can assume their origin is gravitational. A possible gravitational wave detector based on the voltage-to-frequency conversion at a quadrupole Josephson junction device is suggested in Figure \ref{fig2}. In the device, the passing gravitational wave will cause a voltage drop across the quadrupole configuration and a corresponding frequency modulation of the current between the cathode and anode will present itself. There should be a direct correspondence between the amplitude of the passing gravitational wave and the current's frequency. Since such a device can be manufactured in a plane, then manipulating the device's plane in space can yield information on the direction of the source.

Lastly, a cautionary note should be issued when it comes to the design and execution of quantum computing cores from superconducting tunnel junctions acting as qubits. Based on the discussion above, if the electrical scheme of operation of these qubits has the quadrupole design presented in Figure \ref{fig2}, undesirable modulations due to interaction with the gravitational field might be present and can further deteriorate the coherence times and fault tolerance.

\section{The current-voltage characteristics of the quadrupole configuration}

The current-voltage (I-V) characteristics of a superconducting tunnel junction are highly nonlinear and non-trivial. The current density flowing through a junction changes not only in response to changes in voltage $V(t)$ but is also affected by the presence of a vector potential $\vec{A}$. This relationship is often governed by 
\begin{eqnarray}
j_1=j_0 \sin \left[ \delta_1 + \frac{q}{\hbar} \int V dt + \frac{2q}{\hbar} \int \vec{A} \cdot d \vec{r} \right].
\end{eqnarray} 
A single junction makes no distinction between these two modes of phase modulation.

Now we turn our attention to the quadrupole configuration. 
Here we make a projection along one of the axis (the horizontal one) of the device depicted in Figure \ref{fig2}. Since there are two currents flowing in opposite directions (L- for the left-hand side and R for the right-hand side) from the center point to the ground, the total current density is the result of the interference between the two:

\begin{eqnarray}
\nonumber	j_x&=&j_R+j_L\\
\nonumber	&=& 2 j_0 \sin \left[\frac{\delta_R + \delta_L}{2} + \frac{q}{\hbar} \int V dt\right] \\
	&&\qquad \times\cos \left[ \frac{\delta_R - \delta_L}{2} + \frac{2q}{\hbar} \int_R A_x d x\right].
\end{eqnarray} 
The simplification comes out of the relation $\int_R A_x d x = -\int_L A_x d x$ which holds along each axis and is due to the symmetry of the circuit design.

Note the main benefit of the quadrupole circuit configuration: the dependence of the phase of the current on the voltage and the vector potential is now split. Instead of having an oscillation dependent on the vector potential, now the vector potential modulates the amplitude thus producing an envelope for the oscillation determined by the voltage evolution. The vector potential dependence is also restricted to one dimension along each of the arms thus making the device insensitive to the presence of magnetic fields per se. The oscillation of the current through a Superconducting QUantum Interference Device (SQUID) is very sensitive to extremely small magnetic flux, hence its use as a magnetometer. For the quadrupole configuration only an open line integral over the vector potential sets up the amplitude modulation. This configuration is no longer sensitive to magnetic flux and should be operable in noisy environments. 

In effect we have a device which converts voltage to frequency and therefore it can be used in measuring the gravitational fields. The connection is explored in \cite{VA2} and reduces to an effective voltage drop $V_{eff}$ produced by the three-dimensional Ricci scalar curvature $R_{3d}$ of space-time 
\begin{eqnarray}
V_{eff}=\alpha\frac{\hbar^2}{2qm}R_{3d},
\end{eqnarray} 
where $\alpha$ is a numerical factor of the order of $1/6$ to $1/12$. This voltage drop is very small $\approx 10^{-22} R_{3d}$ and can serve as a perturbation only.

In the presence of an applied DC voltage and vanishing initial phases, the output current-voltage characteristic of the device for a non-vanishing space-time curvature becomes
\begin{eqnarray}
	\nonumber	j_x&=&2 j_0 \cos \left[  \frac{2q}{\hbar} \int_R A_x d x\right] \sin\left[\omega_0 \left(1+\frac{\omega_g}{\omega_0}\right)t\right],
\end{eqnarray} 
where $\omega_0={q V}/{\hbar}$, $\omega_g= {\alpha\hbar R_{3d}}/{2m}$ and $\omega_g \ll \omega_0$. Notably, the gravitationally affected signal becomes phase shifted with respect to the DC driven one. The phase shift $\omega_g t$ grows larger, therefore detectable, with time. The measurement can be carried out by comparing the oscillation at the beginning and at a later stage of a continuous run. Conversely, 
\begin{eqnarray}
	j_x& \propto& \omega_g t \cos\left(\omega_0 t\right) + \sin\left(\omega_0 t \right),
\end{eqnarray} 
i.e., for short measuring intervals the current will exhibit an oscillation within a linear envelope determined by $\omega_g t$, and the curvature of space-time can be obtained from the slope of the linear envelope.

\section{Discussion}

The first direct observation of gravitational waves was made in 2015. The effects when measured on Earth are very small, around 1 part in $10^{22}$.
The intriguing idea of our proposal is to use a quantum device for the conversion of electrical power to gravitational wave
emission and possibly detection, where the smallness of $G \times c^{-5}$ can be compensated by $e^4 \times \hbar^{-6}$. This paper goes quite beyond \cite{VA} in the sense that here we solve the nonlinear superconducting junction dynamics exactly and go beyond the interpretations based on the constancy of the supercurrent density. The oscillator interpretation in section \ref{sec.Smallpar} is also revealing since it unambiguously points to a resonant effect coupling the charge distribution to the mass distribution across the junction. 

Interestingly, a dumbbell turned around in the lab also emits gravitational waves with the power of around $10^{-17}$ nW, the Earth revolving around the Sun emits few times $10^{2}$ W and Jupiter a couple of kW. Because of the weak detection efficiency of present detectors, new paradigms are needed. A question therefore arises: Is it possible to construct a resonant device that is able to detect the proposed waves from Josephson devices and weak mechanical sources. Probably not, because the detection is suppressed by the same factor $G \times c^{-5}$. However, let us suppose that in superconducting tunneling it is the bare mass that tunnels and sloshes back and forth. The exact value is hard to evaluate, but let us suppose it is a single digit percent of the mass of the condensate, which is approximately $10^{-4}$ of the mass of the electron content, that is $10^{-6}$ from the $10^{28}$ electron masses, or a nanogram sloshing at 480 THz/V. How does this mass couple to the gravitational wave? For the emission: it is the standard gravitational coupling via the Einstein constant, whose smallness is counteracted by the frequency of oscillation to the power of six, which is quite favourable. For the detection: the gravitational wave shifts the mass a bit, one part in $10^{-22}$, which is the mass in the superconductor that is associated with the charge of the pairs. The redistribution of charges produces a voltage drop accross the tunnel junction and therefore oscillation. The junction is a voltage-to-frequency converter.

Suppose the gravitational wave shifts the masses (ergo charges) $10^{-22}$ m (smaller than a nucleus) in a quadrupole fashion across one arm of the device (single junction), which say
measures $10^{-3}$ m and induces an electric dipole moment $p=10^{-41}$ C.m (10 orders of magnitude less that of the water molecule) which accross 1 mm would produce an electric field and a corresponding voltage drop of $10^{-24}$ V, which is indeed, very small. The large Josephson constant would lead to a current
modulation on the order of $10^{-10}$Hz.

Yet, there may be a way out of this predicament. Suppose the device is extended to include multiple tunnel junctions on each arm like the Josephson voltage standard. In this way the voltage drop can be increased to a measurable value. The new device configuration would resemble the scheme
$$
\langle - {\rm X}-{\rm X}-...-{\rm X}-|-{\rm X}-...-{\rm X}-{\rm X}- \rangle
$$
and may include large number (thousands or millions) of individual junctions. 

Even if the smallness of the detection effect is circumvented by an extended device, the shot and thermal noise from the electrons can swamp any possible gravitational wave detection. Noise sets the limits of performance for all electronic devices. For superconducting tunnel junctions, models exist and reasonably agree with experimental results obtained on relevant tunnel barriers \cite{noise}.

The three types of noise regimes in superconducting junctions have the following properties: (i) $1/f^{\alpha}$ noise (flicker noise) is not fully understood but is believed to be associated with charge traps in the imperfect oxide barrier. It has a marked peak at low frequencies which is unfavourable when it comes to the suggested detection use, yet is amenable by changing the manufacturing technique. ( ii) At intermediate frequencies, it is the white noise (Nyquist-Johnson shot noise) that dominates and grows linearly with frequency. It arises as a form of thermal noise which at 1K starts at around 25 GHz. It is not a primary concern in the detection scheme. Moreover, the coherent quantum state, that is the supercurrent is hardly affected by it. It is only the interaction with the environment, the thermal bath, that drives it. (iii) At very high frequencies it is the quantum noise that reigns. However, the system response is not instantaneous (unlike the response to the gravitational field) and is limited by the RC or L/R characteristic time $\tau$, that is, there is a characteristic cut off frequency of $f_{\rm cut-off} \propto 1/\tau$. This type of noise is rarely seen. In the case of emission, the suggested device may include this type of noise, which is random and does not lead to quadrupole mass distribution, i.e. gravitational wave generation. We, therefore, consider its effect on emission as negligible. As a result, of all types of noises, we consider the flicker noise as the most detrimental to the detection of gravitational waves using a quadrupole superconducting tunnel junction device. It seems that the only solution to the prevention of flicker noise swamping the detection of gravitational waves is advanced manufacturing techniques. Their discussion is beyond the scope of the present paper.

Finally, there is one last scheme to make a measurement of the proof-of-principle: create an emitter and a
detector combo (identical devices). The driving voltage in the emitter should produce a voltage drop in the detector of the same order of magnitude (provided nearby). One has to make sure they are EM decoupled and prove
coincidence to verify the emitting/detecting properties of the proposed device. Here we assume that the generated GW will have the frequency of the oscillation in the emitter condensate. The detector's charges will be driven by a GW with the same frequency (and an unknown amplitude), thereby producing an AC Josephson effect. By virtue of the oscillator model, the detector will  be driven resonantly.

\section{Conclusions}

In consclusion, we would like to point out the main result of this paper - the charge density across a superconducting tunnel junction changes with time, i.e. it is a rapidly oscillating function. The characteristic frequency of this oscillation is proportional to the coupling energy over Planck's constant, that is $\propto K_0 /\hbar $. Since the condensate's pairs are particles with a non-vanishing effective mass, at the superconducting tunnel junction we have a condensed matter set up where rapid mass oscillations can arise. As a result, in an emitting device (two junctions on a line)  electrical power can be converted to gravitational wave emission. A detecting device has a quadrupole configuration of four junctions, where the strain in space-time is converted to a frequency of oscillation, that is a strain-to-frequency conversion takes place.  We believe, these compelling cases can inspire an experimental verification attempt and also open the way to a table-top experimental exploration of gravitation.

One of the main outcomes from the exact solution to the tunneling problem for a superconducting condensate is the realization that the mass dynamics is following the charge dynamics practically identically. Gravitational wave emission/detection research is concentrated around time-dependent quadrupole mass moment which we have demonstrated can be produced in the superconductor context as well. The absence of suggestions to use superconducting tunnel junctions to generate the proper quadrupole mass moment has to be associated with the deeply rooted belief that the approximate (or time-averaged) constant density solution is ``exact". The paper presents evidence for the necessity to move away from the ``constant density" interpretation of the condensate's dynamics and enter the realm of rapid charge/mass oscillations which may find applications outside gravitational physics as well.  

The authors wrote the paper without outside assistance or influence. Useful discussion with Prof. Ralf Sch\"utzhold is acknowledged. Author contributions are as follows: V. A. devised and wrote the initial manuscript with valuable inputs, discussion, comments and edits from A. S. The work of A. S. at Los Alamos National Laboratory was carried out under the auspices of the U.S. DOE and NNSA under Contract No. DEAC52-06NA25396.

No new data were created or analysed in this study.

\end{document}